\documentclass[aps,pra,preprint,a4paper,groupedaddress,showpacs,showkeys]{revtex4-1}
\usepackage{graphicx, subfigure}
\usepackage{amsthm,amssymb,amsmath,amsfonts,bm}

\DeclareMathOperator{\sign}{ sgn}
\begin{document}

\title{
Mass-ratio condition for non-binding of three two-component particles 
with contact interactions 
} 

\author{O.~I.~Kartavtsev} 
\email{oik@nusun.jinr.ru} 

\author{A.~V.~Malykh} 
\email{maw@theor.jinr.ru} 

\affiliation{Joint Institute for Nuclear Research, Dubna, 141980, Russia} 
	
\date{\today}

\begin{abstract} 

Binding of two heavy fermions interacting with a light particle via the contact 
interaction is possible only for sufficiently large heavy-light mass ratio. 
In this work, the two-variable inequality is derived to determine 
a specific value $ \mu^* $ providing that there are no three-body bound states 
for the mass ratio smaller than $ \mu^* $. 
The value $ \mu^* = 5.26 $ is obtained by analyzing this inequality 
for a total angular momentum and parity $ L^P = 1^- $. 
For other $ L^P $ sectors, the specific mass-ratio values providing 
an absence of the three-body bound states are found in a similar way. 
For generality, the method is extended to determine corresponding mass-ratio 
values for the system consisting of two identical bosons and a distinct 
particle for different $ L^P $ ($ L > 0 $) sectors. 

\end{abstract}

\maketitle

\section{Introduction} 
\label{Introduction}

In the recent years, few-body dynamics of multi-component ultra-cold quantum 
gases has attracted much attention. 
A particular form of the short-range interaction between particles becomes 
insignificant in the low-energy limit and the zero-range model provides 
the universal description. 
A particular important is the two-component three-body system with zero-range 
interaction, which has been investigated, e.~g., 
in~\cite{Petrov03,Kartavtsev07,Levinsen09,Helfrich10,Endo11,Castin11,Safavi-Naini13,Jag14,Kartavtsev16,Becker18}. 
A single parameter of the zero-range model, e.~g., a two-body scattering length 
$ a $, can be chosen as a scale, thus there is only one essential parameter, 
the mass ratio of different particles. 
One should mention that introduction of the zero-range model in the few-body 
problem could be ambiguous and needs special efforts, which are discussed, 
e.~g., in~\cite{Minlos12,Minlos14,Minlos14a,Correggi15,Becker18,Kartavtsev16}. 

The system of two identical fermions and a distinct particle 
with the zero-range interaction was considered 
in~\cite{Kartavtsev07,Kartavtsev07a,Endo11,Helfrich11,Michelangeli13,Kartavtsev16} 
and the three-body bound states were found for the mass ratio exceeding some 
critical value. 
Besides these numerical results, it is of interest to find the mass-ratio value 
$ \mu^* $, below which there are no bound states. 
In this work, the values $ \mu^* $ are determined for different sectors of 
a total angular momenta $ L $ and parity $ P $. 
In addition, the non-binding conditions for the system of two bosons 
and a distinct particle are also derived for $ L > 0 $. 
Notice that the value $ \mu^* = 2.617 $ for the fermionic system was obtained 
in~\cite{Becker18} by analyzing the momentum-space integral equation in 
the foremost sector of $ L^P = 1^-$. 

\section{Formulation}
\label{Formulation}

Consider a particle $ 1 $ of mass $ m_1 $ interacting with two identical 
particles $ 2 $ and $ 3 $ of masses $ m_2 = m_3 = m $. 
In the framework of zero-range model, the identical fermions do not interact 
to each other and the same is assumed for generality if the identical particles 
are bosons. 
The zero-range interaction in pairs ($ 1 $ - $ 2 $) and ($ 1 $ - $ 3 $) is 
completely determined by a single parameter, the two-body scattering length $ a $. 
In the center-of-mass frame, one defines the scaled Jacobi variables as 
$ {\mathbf x} = \displaystyle\sqrt{2 \mu }\left( {\mathbf r}_2 - 
{\mathbf r}_1 \right) $ and 
$ {\mathbf y} = \displaystyle\sqrt{2 \tilde \mu} \left({\mathbf r}_3 - 
\dfrac{m_1 {\mathbf r}_1 + m {\mathbf r}_2}{m_1 + m} \right) $, where 
$ {\mathbf r}_i $ is a position vector of $ i $-th particle and 
the reduced masses are denoted by $ \mu = \dfrac{m m_1}{m + m_1} $ and 
$ \tilde \mu = \dfrac{m (m + m_1)}{m_1 + 2m} $. 
The units are chosen by the condition $ \hbar = \vert a \vert = 2 \mu = 1 $, which gives 
the unit two-body binding energy $ \varepsilon_2 = 1 $ for $ a > 0 $. 
The three-body Hamiltonian is a sum of the kinetic energy, which is the minus 
six-dimensional Laplace operator, and the potential energy expressed by 
the boundary conditions imposed at zero distance between the interacting 
particles, 
\begin{equation}
\label{bound1}
\lim_{r \rightarrow 0}\frac{\partial \ln (r\Psi)} {\partial r} = - \mathrm{sign} (a) \ , 
\end{equation} 
where $ r $ denotes either $ \vert{\mathbf r}_1 - {\mathbf r}_2\vert $ or 
$ \vert{\mathbf r}_1 - {\mathbf r}_3\vert $. 
The problem formally depends on a single parameter, the mass ratio $ m/m_1 $, 
alternatively, the kinematic angle $ \omega $ defined by $ \sin \omega = 1/(1 + m_1/m) $ will be used for convenience. 

Total angular momentum $ L $, its projection $ M $ and parity $ {\mathrm P} $  
are conserved quantum numbers, which label the solutions. 
As zero-range interaction acts in the $ s $-wave, it is sufficient to 
consider only the case $ {\mathrm P} = (-)^L $. 
In addition, it is suitable to introduce $ {\mathcal P}_{\mathrm s} $, 
the permutation operator of particles $ 2 $ and $ 3 $, whose eigenvalues 
$ {\mathrm P}_{\mathrm s} = \mp 1 $ designate whether the identical particles 
are fermions or bosons. 

It is convenient to introduce a hyper-radius $ \rho $ and hyper-angles 
$ \{ \alpha, \hat{\mathbf x}, \hat{\mathbf y} \} $ by 
$ x = \rho \cos \alpha $, $ y = \rho \sin \alpha $, $ \hat{\mathbf x} = 
{\mathbf x}/x $, and $ \hat{\mathbf y} = {\mathbf y}/y $. 
In these variables the Hamiltonian is expressed by
\begin{equation}
\label{H}
\displaystyle {\mathrm H} = -\dfrac{1}{\rho^5}\dfrac{\partial}{\partial\rho}
\left( \rho^5 \dfrac{\partial}{\partial \rho } \right) +
\dfrac{\Delta_\Omega }{\rho^2} 
\end{equation} 
supplemented by the boundary conditions, which follow from~(\ref{bound1}). 
In Eq.~(\ref{H}) $ \Delta_\Omega $ denotes the Laplace operator 
on a hyper-sphere, whose explicit form can be found,~e.~g., 
in~\cite{Kartavtsev07,Kartavtsev16,Kartavtsev19}. 
One defines an auxiliary eigenvalue problem on a hyper-sphere (for fixed 
$ \rho $), 
\begin{equation}
\label{eqonhypw}
\left( \Delta_\Omega + \gamma^2(\rho ) - 4 \right) 
\Phi( \alpha, \hat{\mathbf x}, \hat{\mathbf y}; \rho ) = 0 ,
\end{equation}
\begin{equation}
\label{bch}
\lim_{\alpha \rightarrow \pi/2} \frac{\partial\log\left[ (\alpha - \pi/2) 
	\Phi (\alpha, \hat{\mathbf x}, \hat{\mathbf y}; \rho) \right] }
{\partial\alpha }  = \rho \, \sign (a) \, ,  
\end{equation} 
whose eigenfunction $ \Phi (\alpha, \hat{\mathbf x}, \hat{\mathbf y}; \rho) $ 
inherits symmetry of the total wave function and will be chosen 
in the form~\cite{Kartavtsev07,Kartavtsev07a,Kartavtsev16} 
\begin{equation}
\label{Phi}
\Phi (\alpha, \hat{\mathbf x}, \hat{\mathbf y}; \rho ) = 
({\mathcal P}_{\mathrm s} + {\mathrm P}_{\mathrm s})
\frac{\varphi_{\gamma}^L(\alpha )}{\sin 2\alpha} Y_{LM}(\hat{\mathbf y}) \, .
\end{equation}
Here $ Y_{LM}(\hat{\mathbf y}) $ is a spherical 
function, the labels $ L $, $ M $, and $ {\mathrm P}_{\mathrm s} $ in 
the left-hand side are suppressed for brevity, and the function 
$ \varphi_{\gamma}^L(\alpha ) $ satisfies the equations 
\begin{subequations}
\label{bc}
\begin{equation} 
	\label{eqonhyp1}
\left[\frac{d^2}{d \alpha^2} - \frac{L(L + 1)}{\sin^2\alpha}
+ \gamma^2\right]\varphi_{\gamma}^L(\alpha) = 0 \ , 
\end{equation} 
\begin{equation}
	\label{bconhyp}
\lim_{\alpha\rightarrow \pi/2}
\left(\frac{d}{d \alpha} - \rho\, \sign (a) \right) 
\varphi_{\gamma}^L(\alpha) = \frac{2 (-)^L {\mathrm P}_{\mathrm s}}
{\sin 2\omega} 	\varphi_{\gamma}^L(\omega) \, ,
\end{equation} 
\begin{equation}
	\label{bconhyp1}
\varphi_{\gamma}^L(0) = 0 \ .  
\end{equation} 
\end{subequations}
Solution of Eqs.~(\ref{bc}) gives 
the transcendental equation~\cite{Kartavtsev19,Kartavtsev16}
\begin{align}
\nonumber
\rho \sign(a)& 
\Gamma\left(\frac{L + \gamma  + 1}{2} \right) 
\Gamma\left(\frac{L - \gamma  + 1}{2} \right)  = 
2 \Gamma\left(\frac{L + \gamma }{2} + 1 \right) 
\Gamma\left(\frac{L - \gamma }{2} + 1 \right)  \\ 
&+{\mathrm P}_{\mathrm s} 
\frac{(-2)^{1 - L} \pi (\sin\omega )^L}{\sin\gamma \pi \cos\omega } 
\left( \frac{1}{\sin \omega }\dfrac{d}{d  \omega }\right)^L  
\frac{\sin\gamma \omega }{\sin\omega } \, , 
\label{transeq}
\end{align} 
which determines an infinite set of eigenvalues $ \gamma_n^2(\rho ) $ and 
corresponding eigenfunctions 
$ \Phi_n (\alpha, \hat{\mathbf x}, \hat{\mathbf y}; \rho) $. 
Using the expansion 
\begin{equation}
\label{Psi}
\displaystyle
\Psi ({\mathbf x}, {\mathbf y}) = \rho^{-5/2} \sum_{n = 1}^{\infty} f_n(\rho) 
\Phi_n(\alpha, \hat{\mathbf x}, \hat{\mathbf y}; \rho) \, ,
\end{equation}
one obtains a system of hyper-radial 
equations~\cite{Kartavtsev07,Kartavtsev16,Kartavtsev19} for the channel 
functions $ f_n(\rho) $. 
In the following analysis, the mass-ratio bound $ \mu^* $ will be found 
by using the one-channel approximation, 
\begin{equation}
\label{system1}
\left[\frac{d^2}{d \rho^2} - \frac{\gamma^2(\rho) - 1/4}{\rho^2} + E \right]
f(\rho)  = 0 \, ,
\end{equation}
where $ \gamma^2(\rho) $ and $ f(\rho) $ denote the lowest eigenvalue and 
corresponding channel function, in addition, 
$ \Phi(\alpha, \hat{\mathbf x}, \hat{\mathbf y}; \rho) $ will be used to denote 
the corresponding eigenfunction on a hyper-sphere.  
One should emphasize an additional approximation made in Eq.~(\ref{system1}), 
namely, the term 
$ \displaystyle \int \left( \dfrac{\partial \Phi }{\partial \rho} \right)^2 
\sin^2 \alpha d \alpha d \hat{\mathbf x} d \hat{\mathbf y} $ is omitted. 

As it was discussed 
in~\cite{Kartavtsev16,Kartavtsev19,Minlos14,Correggi15,Becker18}, 
the formal description of the three-body problem is not sufficient
for the mass ratio exceeding a specific value $ \mu_r $ defined by 
the condition $ \gamma^2(0) = 1 $ from Eq.~(\ref{transeq}). 
For unambiguous definition, it is necessary to impose an additional boundary 
condition, which determine the wave function at the triple collision point 
(for $ \rho \to 0 $). 
Below it will be confirmed that $ \mu^* < \mu_r $, therefore, the problem 
is completely defined by the requirement of square integrability or, 
equivalently, by the boundary condition $ f(\rho) \xrightarrow[\rho \to 0]{} 0 $. 

As follows from~(\ref{transeq}), $ \gamma^2 (\rho) \ge 1 $ for any 
$ m/m_1 < \mu_r $ if either $ a > 0 $ and 
$ {\mathrm P}_{\mathrm s} = (-)^{L + 1} $ or $ a < 0 $, which entails absence 
of bound states. 
Besides, it is trivial that an infinite number of bound states exist for any 
$ m/m_1 $ in the case $ L^P = 0^+ $ for two identical bosons and a distinct 
particle ($ {\mathrm P}_{\mathrm s} = 1 $). 
Thus, it remains to determine the value $ \mu^* $ only for the positive 
scattering length ($ a > 0 $) and $ {\mathrm P}_{\mathrm s} = (-)^{L} $, 
i.~e., for odd $ L $ and $ P $ (even $ L > 0 $ and $ P $) if the identical 
particles are fermions (bosons). 

\section{Non-binding condition} 
\label{Non-binding} 


For determination of the specific value $ \mu^* $ it is sufficient to construct 
the lower bound $ E_{LB} $ of the exact three-body energy $ E $ and prove that 
$ E_{LB} $ exceeds the two-body threshold for $ m/m_1 \le \mu^* $. 
This will be done by the following three steps. 

\paragraph*{\rm{1.}} 
The lower bound $ E_{LB} $ for the ground state energy $ E $ will be obtained 
by solving Eq.~(\ref{system1}). 
It follows from the general statement on the lower bound of energy for any 
Hamiltonian separated into two parts
\begin{equation}
\label{Ham}
{\mathrm h} = {\mathrm T}_1 + V_1 (\xi ) + {\mathrm T}_2 + V_2 (\xi, \eta ) \, , 
\end{equation}
where $ \xi $ and $ \eta $ denote the sets of "slow" and "fast" variables. 
The kinetic energies $ {\mathrm T}_1 $ and $ {\mathrm T}_2 $ depend on $ \xi $ 
and $ \eta $, respectively. 
Denoting the lowest eigenvalue of $ {\mathrm T}_2 + V_2 (\xi, \eta ) $ by 
$ \varepsilon (\xi ) $, one obtains that the lowest eigenvalue $ E_{LB} $ of 
$ {\mathrm T}_1 + V_1 (\xi ) + \varepsilon (\xi ) $ is the lower bound for 
all eigenvalues of the initial Hamiltonian $ {\mathrm h} $,~i.e., 
$ E_{LB} \le E $. 

To sketch a simple proof, one should notice that any operator is bounded 
from below by its least eigenvalue, i.~e., 
\begin{equation}
\label{T2V2eq}
{\mathrm T}_2 + V_2 (\xi, \eta ) \ge \varepsilon (\xi ) \, , 
\end{equation}
where the operator inequality $ {\mathrm A} \ge {\mathrm B} $ means that $ \left\langle \phi \vert{\mathrm A}\vert 
\phi \right\rangle \ge \left\langle \phi \vert{\mathrm B}\vert \phi \right\rangle $ 
for any $ \phi $. 
Using that $ {\mathrm A} + {\mathrm B} \ge {\mathrm A} + {\mathrm C} $, if  $ {\mathrm B} \ge {\mathrm C} $, one obtains 
$ {\mathrm h} \ge {\mathrm T}_1 + V_1  + \varepsilon (\xi ) $ and 
$ E \ge E_{LB} $. 

This lower bound for the ground state energy was multiply discussed in 
the literature, e.~g., this line of proof was carried out for the adiabatic 
description of molecules~\cite{Brattsev65, Epstein66}, the $ N $-body problem 
within the hyper-spherical framework~\cite{Coelho91}, and the hydrogen atom 
in magnetic field~\cite{Starace79}. 

These arguments can be applied to the problem under consideration by using 
the hyper-radius $ \rho $ as a "slow" variable and the hyper-angles 
$ \{ \alpha, \hat{\mathbf x}, \hat{\mathbf y} \} $ as "fast" variables. 
Then the kinetic-energy operator on the hyper-radius is taken as 
$ {\mathrm T}_1 $, the potential $ V_1 = 0 $, Eqs.~(\ref{eqonhypw}) 
and~(\ref{bch}) determine $ {\mathrm T}_2 + V_2 $, and 
$ \dfrac{\gamma^2}{\rho^2} $ essentially corresponds to $ \varepsilon (\rho ) $. 
Thus the eigenvalue equation for $ {\mathrm T}_1 + V_1 + \varepsilon (\rho ) $ is 
equivalent to Eq.~(\ref{system1}), which solution provides the lower bound for 
the exact energy of the original problem. 

\paragraph*{\rm{2.}} 
Introduce the reference Hamiltonian 
$ {\mathrm h}_r = -\dfrac{d^2}{d x^2} - \dfrac{1}{4 x^2} $ under the requirement 
that only the functions $ \varphi (x) $ satisfying the condition 
$ \varphi \displaystyle  \xrightarrow[x \to 0]{} x^{1/2}[1 + O(x)] $ 
are admitted. 
It is well-known that $ {\mathrm h}_r $ is non-negative, i.~e., 
$ \left\langle \varphi \vert {\mathrm h}_r \vert \varphi \right\rangle \ge 0 $ for any 
$ \varphi $, as $ {\mathrm h}_r $ is a radial part of the two-dimensional 
kinetic-energy operator. 
Thus, $ {\mathrm h}_r $ possesses only continuous spectrum and there are no 
bound states for any operator $ \Tilde{\mathrm h} $ if 
$ \Tilde{\mathrm h} \ge {\mathrm h}_r $. 
Nevertheless, a bound state of $ {\mathrm h}_r + V(x) $ arises for an arbitrarily 
small $ V(x) $ provided $ \int V(x) x dx \le 0 $ and it is known that 
the bound-state energy is exponentially small for $ V(x) \to 0 $~\cite{Simon76}. 

\paragraph*{\rm{3.}} 
Determining whether the operator in Eq.~(\ref{system1}) exceeds the reference 
Hamiltonian $ {\mathrm h}_r $, one obtains the condition on $ \gamma^2(\rho ) $ 
that ensures non-existence of bound states. 
Taking into account that the two-body threshold $ \epsilon_2 = -1 $ and 
$ \gamma^2(\rho )/\rho^2 $ tends to this value for $ \rho \to \infty $, one 
finds that the bound states do not exist if $ \gamma^2(\rho )/\rho^2 \ge -1 $, 
i.~e., if $ \gamma^2/\rho^2 $ exceeds the threshold $ \epsilon_2 = -1 $. 
Denoting $ \gamma = i \varkappa $, one obtains the final result  
\begin{equation} 
\label{cond}
\rho(i\varkappa) - \varkappa \ge 0\, ,
\end{equation}
where the function $ \rho(\gamma ) $ is given by~(\ref{transeq}). 
For $ P = P_s = (-1)^L $ and $ a > 0 $ the condition~(\ref{cond}) takes 
the form $ \ B_L(\varkappa, \omega) \ge 0 $, where  
\begin{align}
\nonumber
B&_L(\varkappa,\omega) \equiv  
{2 \Gamma\left(\frac{L + i \varkappa }{2} + 1 \right) 
\Gamma\left(\frac{L - i \varkappa }{2} + 1 \right) } - \frac{2^{1 - L} \pi (\sin\omega )^L}{\sinh \varkappa \pi \cos\omega } \\  
&\times \left( \frac{1}{\sin \omega }\dfrac{d}{d  \omega }\right)^L 
\frac{\sinh  \varkappa \omega }{\sin\omega } - 
 \varkappa\,{\Gamma\left(\frac{L + i \varkappa  + 1}{2} \right)
\Gamma\left(\frac{L - i \varkappa  + 1}{2} \right)}.
\label{BL}
\end{align} 
Finally, the problem reduces to finding the value $ \omega^* $ providing that 
the condition~(\ref{BL}) is satisfied for any 
$ 0 < \omega \le \omega^* < \pi/2 $ and $ \varkappa > 0 $, then $ \mu^* $ follows 
from the relation $ \sin\omega^* = \mu^*/(1 +\mu^*) $.

\subsection{Fermionic system in the sector $ L^P = 1^- $} 
\label{L1}

With increasing the mass ratio, the first three-body bound state in 
the system of identical fermions and a distinct particle arises in the sector 
of angular momentum and parity $ L^P = 1^- $. 
In this case the mass-ratio condition for non-binding, i.~e., 
the inequality $ \ B_1(\varkappa, \omega) \ge 0 $, takes the form, 
\begin{equation}
\label{lowboundL1}
\hskip -.7cm
F(\varkappa) - G(\varkappa , \omega ) \ge 0 ,
\end{equation}
where $ F(\varkappa) = \displaystyle \left( \varkappa^2 + 1 \right) 
\sinh \frac{\varkappa \pi}{2} - \varkappa^2 \cosh \frac{\varkappa\pi}{2} $
and $ G(\varkappa,\omega) = \displaystyle 2\varkappa\frac{\cosh\varkappa\omega}{\sin2\omega} - 
\frac{\sinh \varkappa\omega}{\sin^2 \omega} $. 

Firstly, prove that $ \ B_1(\varkappa, \omega) \ $ is monotonically decreasing 
function of $\omega $ ($ 0 < \omega < \pi/2 $) for any $ \varkappa > 0 $. 
The condition 
$ \dfrac{\partial B_1(\varkappa, \omega )}{\partial\omega } \le 0 $ 
is equivalent to 
$ \dfrac{\partial G(\varkappa, \omega )}{\partial\omega } \ge 0 $, which 
explicitly gives  
\begin{equation}
\hspace{-9mm}
\left( \varkappa^2 \tan \omega + 2 \cot \omega \right) \tanh \varkappa \omega + 
\varkappa\left(\tan^2 \omega - 2 \right) \ge 0 . 
\label{ineq1} 
\end{equation}
The inequality~(\ref{ineq1}) is evidently fulfilled for 
$ \tan^2 \omega \ge 2 $.  
To proceed further, after simple transformations the inequality is written as 
\begin{equation} 
\label{ineq2}
\left[ 4 + \varkappa^2 (\varkappa^2 + 4) z^2 - \varkappa^2 z^3 \right] 
\sinh^2\varkappa\omega - \varkappa^2 z (z - 2)^2 \ge 0 \, ,
\end{equation}
where $ z = \tan^2 \omega $ is used for brevity. 
Using the inequality $ \sinh^2 \varkappa \omega \ge 
\varkappa^2 \sin^2 \omega \equiv \varkappa^2 z/(1 + z) $ in Eq.~(\ref{ineq2}), 
one comes to a simple result
\begin{equation} 
\label{ineq3}
\varkappa^2 + 3 - z \ge 0 \, , 
\end{equation}
which is fulfilled for any $ \varkappa $ if $ z \equiv \tan^2 \omega \le 3 $. 
This completes the proof that $ \dfrac{\partial B_1}{\partial\omega } \le 0 $, 
therefore, the implicit condition $ B_1(\varkappa, \omega ) = 0 $ determines a 
single-valued function $ \omega_{0}(\varkappa ) $.
At last, if one finds 
\begin{equation}
\omega^* = \min \omega_0(\varkappa ), \quad \quad 0 \le \varkappa < \infty 
\end{equation} 
and corresponding mass ratio $ \mu^* $, it provides absence of bound states 
for any $ \omega \le \omega^* $ (consequently, for $ m/m_1 \le \mu^* $). 

\begin{figure}[t!] 
\centering
\includegraphics[width=0.5\textwidth]{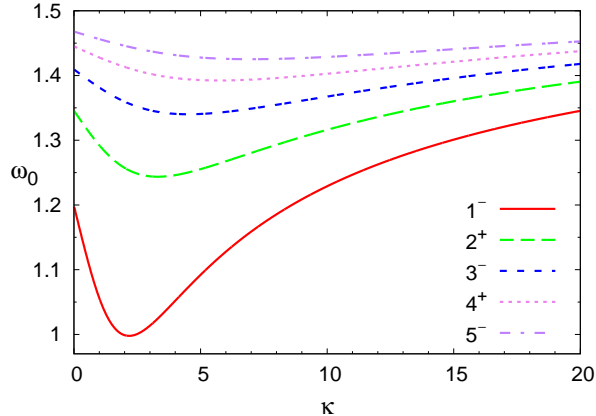} 
\caption{Dependence $ \omega_{0}(\varkappa ) $ 
for different $ L^P $ states. 
Odd (even) total angular momentum $ L $ and parity $ P $ correspond 
to the system containing two identical fermions (bosons) and a distinct 
particle.} \label{fig1}
\end{figure}

The function $ \omega_0 (\varkappa ) $, as shown in Figure~\ref{fig1}, has one 
minimum at $ \varkappa^* $ and its value 
$ \omega^* = \omega_{0}(\varkappa^* ) $ determines the mass ratio $ \mu^* $. 

Numerical values $ \varkappa^* $, $ \omega^* $, and $ \mu^* $ are presented in 
Table~\ref{tab1}. 
Recall that the condition $ m/m_1 \le \mu^* $ in the sector $ L^P = 1^-$ 
provides absence of any bound states in this system. 
Now it is possible to confirm an assumption of Section~\ref{Formulation}, 
namely, the condition $ \mu^* < \mu_r $ is valid as the critical mass ratio 
$ \mu_r \approx 8.6185769247 $ for $ L^P = 1^- $. 
This means that, as discussed in Section~\ref{Formulation}, the three-body 
Hamiltonian is completely defined by the requirement of square integrability or 
one can simply use the zero boundary condition at the triple collision point. 
\begin{table}[h]
\begin{center}
\caption{Mass-ratio values $ \mu^* $ and corresponding $ \varkappa^* $ 
and $ \omega^* $ in different $ L^P $ sectors. 
Odd (even) $ L $ and $ P $ correspond to the system containing two identical 
fermions (bosons) and a distinct particle. 
$ \mu_B $ is the mass ratio, at which the first bound state appears, 
found by numerical calculations~\cite{Kartavtsev07,Kartavtsev19}.} 
\label{tab1}
\begin{tabular}{@{}ccccc@{}}
$ L^P $ & $ \varkappa^* $ & $ \omega^*$ & $ \mu^* $ & $ \mu_B $ \\
\hline
$ 1^- $ & 2.17701 & 0.997755 &  5.26002 & 8.17259 \\
$ 2^+ $ & 3.30822 & 1.243618 & 17.85119 & 22.6369 \\
$ 3^- $ & 4.51245 & 1.340135 & 36.75782 & 43.3951 \\
$ 4^+ $ & 5.74050 & 1.392347 & 61.97274 & 70.457  \\
$ 5^- $ & 6.97890 & 1.425184 & 93.49356 & 103.823 \\
\end{tabular}
\end{center}
\end{table} 

\subsection{Angular momenta $ L \ge 2 $}
\label{Lany}

Besides the general non-binding condition, it is of interest to derive also 
the corresponding conditions in any $ L^P $ sectors. 
As discussed previously, one should consider odd (even) $ L $ and $ P $ for 
the system containing two identical fermions (bosons) and a distinct particle. 
Using the described approach, the values $ \mu^* $ will be determined below for 
$ L \le 5 $. 
Analogously to preceding Section~\ref{L1}, one supposes that 
$ B_L(\varkappa, \omega) $ in Eq.~(\ref{BL}) monotonically decrease 
with increasing $ 0 < \omega < \pi/2 $ for any $ \varkappa > 0 $. 
Thus the condition $ B_L(\varkappa, \omega) = 0 $ again determines 
a single-valued function $ \omega_{0}(\varkappa ) $ and finding its global 
minimum $ \omega^* $ provides absence of bound states for any 
$ \omega \le \omega^* $, correspondingly, for $ m/m_1 \le \mu^* $. 

The expressions for $ B_L(\varkappa,\omega) $ become lengthy and difficult 
to handle for higher $ L $. 
In particular, for the three-body system containing two identical bosons in the sector $ L^P = 2^+ $, 
\begin{align}
\nonumber
  B_2(\varkappa,\omega) &= \frac{(1+\varkappa^2)\pi}{2\sinh \varkappa\pi}\left[\frac{\varkappa(\varkappa^2 + 4)}{\varkappa^2 + 1} 
\cosh\frac{\varkappa\pi}{2} - \varkappa\sinh\frac{\varkappa\pi}{2} +\right. \\ 
&+  \left.
3 \frac{\varkappa \cosh\varkappa\omega - \sinh \varkappa\omega \cot\omega}
{(\varkappa^2 + 1) \sin^2 \omega } - 
\frac{2 \sinh \varkappa\omega}{\sin 2\omega} \right] \, .
\label{lowboundL2} 
\end{align}

Numerical calculations reveal that for all $ 0 < L \le 5 $ the functions 
$ \omega_{0}(\varkappa ) $ exhibit one minimum, as shown in Figure~\ref{fig1}. 
The positions of these minima ($ \varkappa^*, \omega^* $) are calculated 
and presented in Table~\ref{tab1} jointly with corresponding values $ \mu^* $. 
All the values $ \mu^* $, $ \omega^* $, and $ \varkappa^* $ increase with 
increasing $ L $, thus reflecting a general trend for $ L $-dependence of 
the critical mass-ratio value $ \mu_B $, at which the first bound state appears. 
For comparison, the result of numerical 
calculations~\cite{Kartavtsev19, Kartavtsev07} of $ \mu_B $ is presented in 
the last column of Table~\ref{tab1}. 
The relative difference $ \dfrac{\mu^* - \mu_B}{\mu_B} $ decreases from 
$ 0.36 $  to $ 0.1 $ for increasing angular momentum from $ L = 1 $ to $ L = 5 $. 
Again, it is possible to confirm an assumption in Section~\ref{Formulation} 
that $ \mu^* < \mu_r $ for considered total angular momenta as $ \mu^* < \mu_B $
and $ \mu_B < \mu_r $~\cite{Kartavtsev16, Kartavtsev19}. 

\section{Conclusion}

Using the one-channel approximation for a system of hyper-radial equations, 
the non-binding condition for three particles is written as 
an inequality for the function of two variables. 
In this way, it was proven that two fermions and a distinct particle are 
not bound for any mass ratio below $ \mu^* = 5.26 $. 
This bound is sufficiently close to the result of numerical 
calculations~\cite{Kartavtsev07,Kartavtsev16} for the mass ratio  
$ \mu_B \approx 8.17259 $, at which the first bound state arises. 
This non-binding condition on the mass ratio was obtained by considering 
the states of total angular momentum and parity $ L^P = 1^- $ for positive 
two-body scattering length $ a > 0 $. 
So far the lower bound $ \mu^* = 2.617 $ was obtained from analysis of 
the momentum-space integral equations~\cite{Becker18}. 

Furthermore, the same procedure was used to find the non-binding conditions 
also for the states of higher total angular momenta $ L \le 5 $ 
for three-body systems containing either two identical fermions or 
two non-interacting bosons. 
The non-binding conditions was determined for odd (even) $ L $ and $ P $
for  the system containing fermions (bosons). 
As expected, the one-channel approximation works better for higher $ L $ that 
leads to better agreement between $ \mu^*$ and the numerically calculated 
$ \mu_B $, at which the first bound state arises in given $ L^P $ sector. 

One can hope that the described method will be useful for determination of 
the lower bounds in other three-body problems.

\end{document}